\documentclass[10pt,twocolumn,letterpaper]{article}
\usepackage{cvpr}             
\usepackage{graphicx}
\usepackage{amsmath}
\usepackage{amssymb}
\usepackage{booktabs}
\usepackage{tikz}
\usepackage{multirow}
\usepackage{multicol}
\usepackage{float}
\usetikzlibrary{shapes.geometric, arrows}
\usepackage[skip=5pt]{caption} 
\usepackage[breaklinks,colorlinks]{hyperref}
\usepackage[capitalize]{cleveref}
\crefname{section}{Sec.}{Secs.}
\Crefname{section}{Section}{Sections}
\Crefname{table}{Table}{Tables}
\crefname{table}{Tab.}{Tabs.}
\usepackage[style=mla,backend=biber]{biblatex}
\DeclareLanguageMapping{american}{american-apa}
\DeclareUnicodeCharacter{03B2}{\ensuremath{\beta}}
\begin{document}

\title{Machine learning based modeling to predict inhibitors for targets of Alzheimer's Disease}

\author{Mehak Gopal\\
{\tt\small 2021475}
\and
Nalin Arora\\
{\tt\small 2021478}}
\maketitle

\begin{abstract}
Alzheimer's Disease is a chronic neurodegenerative disorder projected to affect 115 million people by 2050, driven by mechanisms like the cholinergic and amyloid hypotheses and insulin signaling disruptions involving key targets such as BACE-1, AChE, and GSK-3$\beta$. Utilizing machine learning (ML), we developed predictive models for inhibitor screening, achieving AUC-ROC scores above 0.9 for all targets. BACE-1 models showed high accuracy (86.63\%) but limited chemical diversity. AChE models exhibited greater chemical diversity and similar performance (AUC-ROC: 92.86\%, Accuracy: 85.20\%), while GSK-3$\beta$ models achieved an AUC-ROC of 91.14\% with the highest proportion of viable drug candidates. These findings highlight the potential of ML in Alzheimer's drug discovery, with AChE and GSK-3$\beta$ emerging as promising targets. 

\end{abstract}

\section{Introduction}
\label{sec:intro}
Alzheimer’s Disease, or AD, is a multifactorial disorder of the brain. It is a chronic progressive neurodegenerative disease that affects more than 10\% of elderly people over 65 years of age. It affects 35 million people all over the world and it can rise up to 115 million by 2050 (\cite{Sandhu2022}). 

This disease is mainly associated with the cholinergic and amyloid hypothesis (\cite{dualinhib}), which leads to increased production of amyloid-$\beta$ protein, which is an essential component of amyloid plaques. These neurotoxins are due to the cleavage of APP, which is controlled by the enzyme \textit{BACE-1}. The cholinergic hypothesis highlights the loss of cholinergic neurotransmission, which is responsible for memory loss and cognitive decline. \textit{AChE} is an enzyme that degrades acetylcholine, a neurotransmitter critical for learning and memory.
AD has also been shown to be linked to diabetes and insulin production. Reduced insulin signaling activates \textit{GSK-3$\beta$}, which leads to the formation of taus and neurofibrillary tangles (\cite{BRIGGS2016247}). Hence, these biological targets can be considered as the top focus for drug research for Alzheimer's. 

With the advent of artificial intelligence, many processes can be streamlined for faster and more accurate inferences. Machine-learning-based modeling provides a cost-effective and efficient strategy for screening and prioritizing inhibitors in drug discovery. Given the complexity of AD pathology and the vast datasets involved, machine learning offers a powerful tool to accelerate the identification of potential therapeutic targets.

\section{Literature Review}
Multiple studies have explored the application of machine learning for predicting inhibitors targeting Alzheimer’s disease. For instance, \cite{noviandy2023qsar} focused on BACE-1 inhibitor activity prediction. This study utilized 2D molecular descriptors combined with a Genetic-Algorithm-based feature selection and ensemble machine learning models, including Random Forest, Gradient Boost, Extra Boost, and AdaBoost. Among these, RF achieved the best performance, with a training accuracy of 98.65\% and a testing accuracy of 82.53\%. Similarly, \cite{Sandhu2022} addressed AChE inhibitor prediction using molecular fingerprints and Random Forest, obtaining a testing accuracy of 85.38\% for their best model.

The enzyme GSK-3$\beta$, another key target in AD pathology, has also been investigated (\cite{ijms242417233}). They modeled a consensus approach using their best models and had an accuracy of 71\% on the test set. They performed virtual screening to identify new inhibitors and were able to identify two, G1 and G4. 

Furthermore, dual prediction models for AChE and BACE-1 inhibitors have been developed. For example, \cite{dualinhib} constructed 2D-QSAR models combining molecular descriptors for spatial, structural, and electronic properties. Using both linear and nonlinear machine learning methods, the study achieved significant predictive accuracy for both inhibitors.

Another notable contribution to BACE-1 inhibitor prediction is the work by \cite{ponzoni2019qsar}, which developed models with an average accuracy of 79\%.

These studies collectively highlight the progress in machine-learning-based modeling for AD drug discovery. However, most of these papers work with small datasets and minimal fingerprints. Their machine learning pipelines aren't rigorous with feature selection and parameter tuning. 

\section{Materials and Methods}
\begin{figure}[ht]
    \centering
    \caption{Pipeline}
    \includegraphics[width=0.5\textwidth]{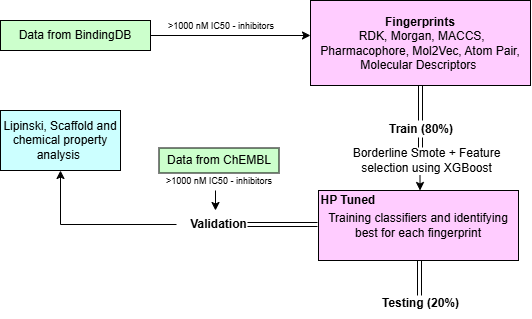}
    \label{fig:pipeline} 
\end{figure}

The entire pipeline has been summarised in figure \ref{fig:pipeline}.

\subsection{Dataset}
We chose three targets for our analysis: BACE-1, AChE, and GSK-3$\beta$. The data was extracted from ChEMBL (\cite{10.1093/nar/gkad1004}) and BindingDB(\cite{bindigdb}). 

ChEMBL curates data from various existing public databases and scientific literature. BindingDB curates US patents and journals that are not covered by other public databases. 

We used BindingDB with their SMILES format for our model development for each molecule. For validation, we used ChEMBL. We made sure there were no overlaps with the training and testing data.
The dataset numbers are expanded on in table \ref{tab:example}.

\begin{table}[ht]
\centering
\caption{Dataset Numbers}
\label{tab:example}
\begin{tabular}{|c|c|c|}
\hline
\textbf{Biological Target} & \multicolumn{2}{|c|} {\textbf{Number of Entries}}\\ \hline & \textbf{BindingDB} & \textbf{ChEMBL} \\ \hline
BACE-1 & 16048 & 6448     \\ \hline
AChE  & 20553 & 7616    \\ \hline
GSK-3$\beta$ & 5742 & 3844      \\ \hline
\end{tabular}
\end{table}

To define active inhibitors, we used $>$1000 nM  $IC_{50}$ (\cite{ponzoni2019qsar}) as a threshold for all the molecules obtained. 

An 80:20 split was done for training and testing. After feature extraction and selection, Borderline SMOTE (\cite{han2005borderline}) was done for the training data to provide the models with a balanced class dataset.
\subsection{Features}
\subsubsection{Feature Extraction}
All the SMILES for each inhibitor were converted to numerical format using multiple methods. 
\begin{itemize}
    \item Molecular descriptors: This was done using RDkit (\cite{RDKit2020}) and Mordred (\cite{mordred}). These gave us a combination of both 2D and 3D descriptors.
    \item Morgan Fingerprints (\cite{Morgan2010}): Based on circular substructures, these hashed fingerprints encode the local atomic environment up to a specified radius (2 for our case).
    \item MACCS Keys (\cite{MACCS2002}): These are 166 pre-defined structural keys that capture molecular features like functional groups and substructures. 
    \item RDK Fingerprints (\cite{RDKit2020}): These are path-based fingerprints that encode molecular connectivity. 
    \item Atom Pair FingerPrints (\cite{AtomPair1985}): They represent pairs of atoms in the molecule and their topological distance, encoding structural and spatial relationships.
    \item Pharmacophore Fingerprints (\cite{Pharmacophore1998}): They encode spatial arrangements of features like hydrogen bond donors, acceptors, and hydrophobic regions. 
    \item Mol2Vec Embeddings (\cite{Mol2Vec2018}): A vectorization technique from the Natural Language Processing domain, the SMILES have been encoded using the Word2Vec algorithm.
\end{itemize}

\subsubsection{Feature Selection}
Considering the vast number of fingerprints and descriptors generated, we used XGBoost as a feature selector for every feature. The data was imputed to deal with NaN values and then standardized. 

For each descriptor, features contributing to the top 90\% of the total importance were selected, and the rest were discarded. This was possible due to the inbuilt feature of "feature\_importances\_" in the XGBoost Classifier.

\subsection{Machine Learning Models}
\begin{figure}[H]
    \centering
    \caption{AUC-ROC for BACE-1}
    \includegraphics[width=0.5\textwidth]{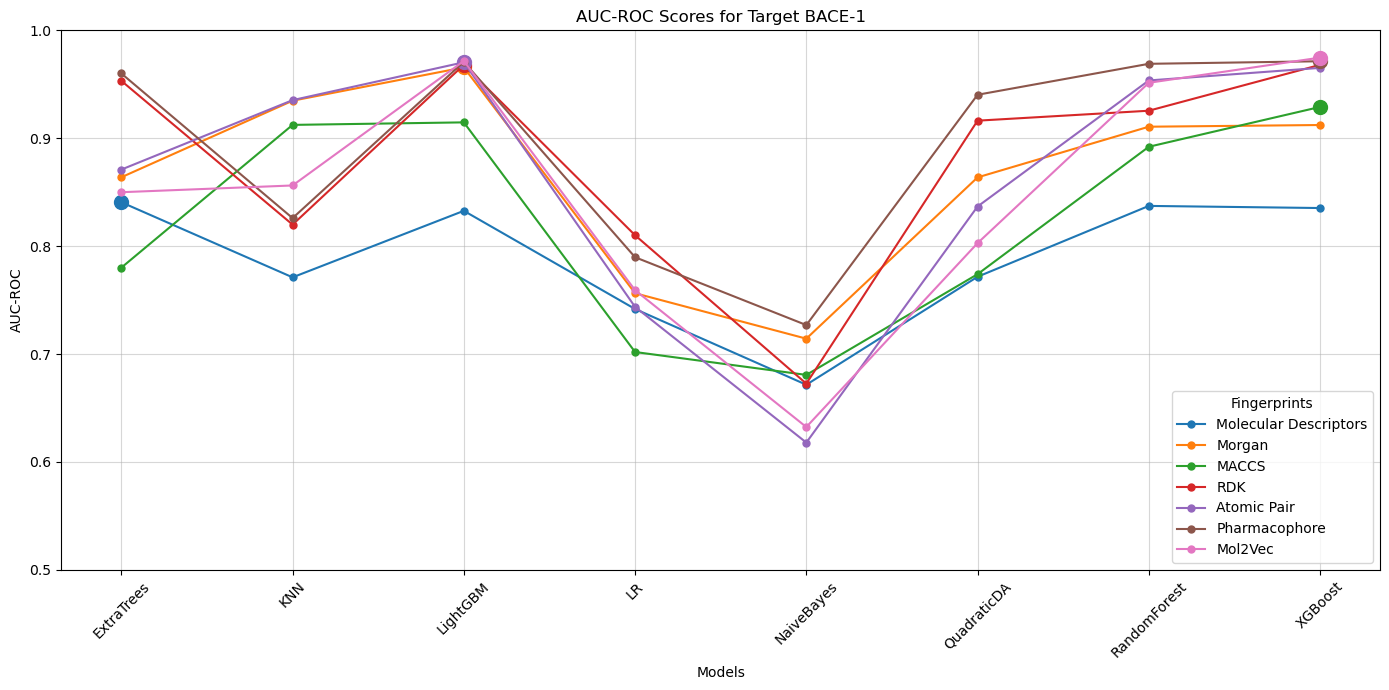} 
    \label{fig:bace_trainign} 
\end{figure}

\begin{figure}[H]
    \centering
    \caption{AUC-ROC for GSK-3$\beta$}
    \includegraphics[width=0.5\textwidth]{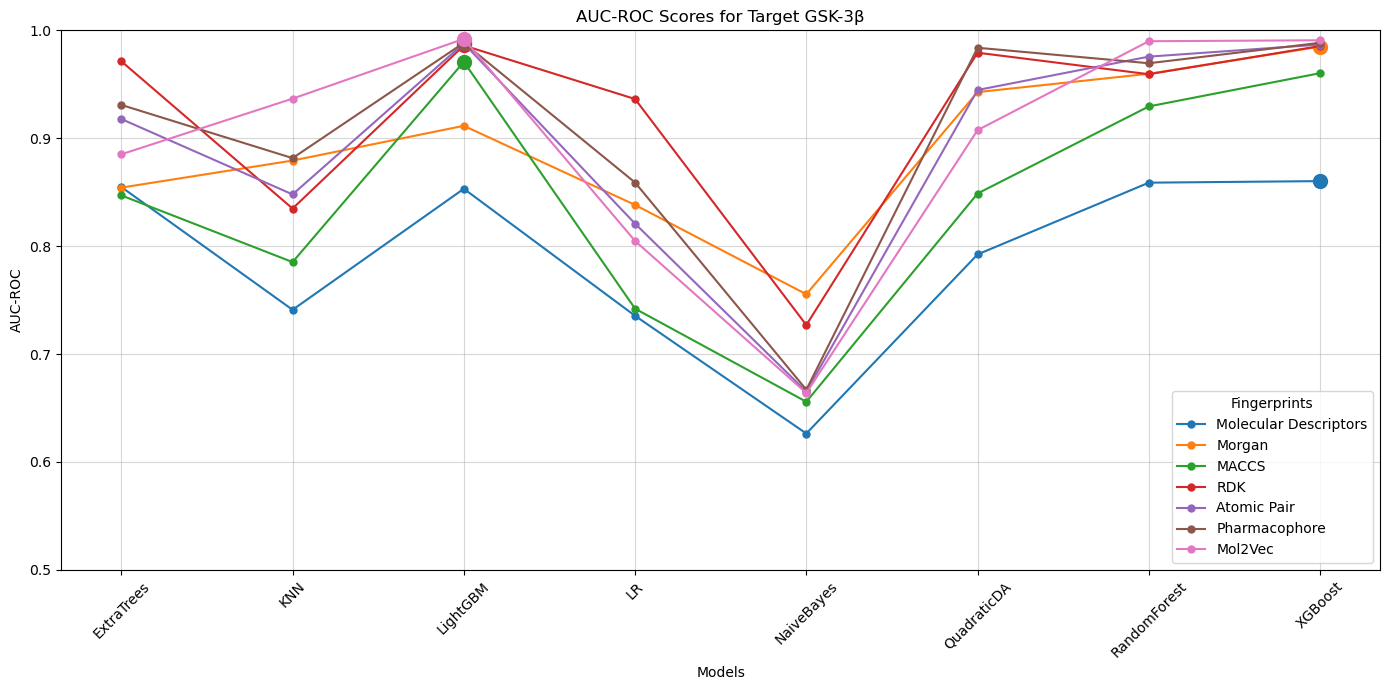} 
    \label{fig:gsk_training} 
\end{figure}

\begin{figure}[H]
    \centering
    \caption{AUC-ROC for AChE}
    \includegraphics[width=0.5\textwidth]{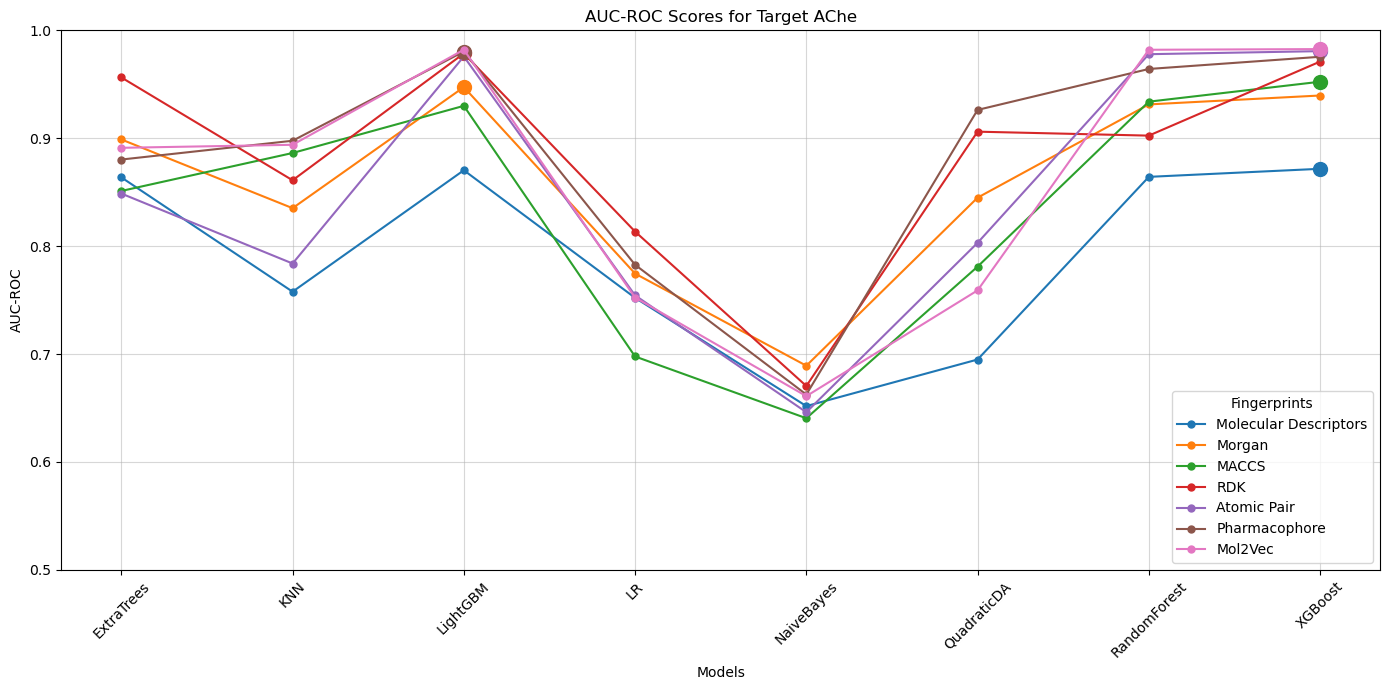} 
    \label{fig:AChE_training} 
\end{figure}

We initially used multiple machine learning models to understand the best classifier. Each ML model was hyperparameter-tuned using Optuna (\cite{akiba2019optuna}). For each fingerprint/descriptor, eight classifier models were tested, i.e., ExtraTrees (ET), KNN, LightGBM, Logistic Regression (LR), NaiveBayes, QuadraticDiscriminantAnalysis (QuadraticDA), Random Forest (RF), and XGBoost. The best model for each target for each fingerprint/descriptor was chosen based on the AUC scores. The best models with the AUC for each target are highlighted in the figures \ref{fig:bace_trainign},\ref{fig:gsk_training} and \ref{fig:AChE_training}. 


As we can see, either XGBoost (XGB) or LightGBM (LGBM) performed the best and were chosen for testing and validation, and their results can be seen in table \ref{tab: trainig_result}.

\begin{table}[H]
\centering
\caption{Best Models and their scores}
\label{tab: trainig_result}
\begin{tabular}{|l|l|l|}
\hline 
\textbf{Fingerprint} & \textbf{Target} & \textbf{Model-Score} \\ \hline
\multirow{3}{*}{Descriptors} 
    & BACE-1         & ET-0.84          \\ \cline{2-3}
    & GSK-3$\beta$         & XGB-0.86    \\ \cline{2-3}
    & AChE        & XGB-0.87         \\ \hline
\multirow{3}{*}{Morgan} 
    & BACE-1         & LGBM-0.96           \\ \cline{2-3}
    &  GSK-3$\beta$         & XGB-0.98     \\ \cline{2-3}
    & AChE           & LGBM-0.94         \\ \hline
\multirow{3}{*}{MACCS} 
    & BACE-1         & XGB-0.92           \\ \cline{2-3}
    &  GSK-3$\beta$          & LGBM-0.97           \\ \cline{2-3}
    & AChE           & XGB-0.95           \\ \hline
\multirow{3}{*}{RDK} 
    & BACE-1         & LGBM-0.96           \\ \cline{2-3}
    &  GSK-3$\beta$          & LGBM-0.98           \\ \cline{2-3}
    & AChE           & LGBM-0.97          \\ \hline
\multirow{3}{*}{Atomic Pair} 
    & BACE-1         & LGBM-0.97           \\ \cline{2-3}
    & GSK-3$\beta$         & LGBM-0.98           \\ \cline{2-3}
    & AChE           & XGB-0.98       \\ \hline
\multirow{3}{*}{Pharmacophore} 
    & BACE-1         & XGB-0.97           \\ \cline{2-3}
    &  GSK-3$\beta$       & LGBM-0.98           \\ \cline{2-3}
    & AChE           & LGBM-0.98           \\ \hline
\multirow{3}{*}{Mol2Vec} 
    & BACE-1         & XGB-0.97           \\ \cline{2-3}
    &  GSK-3$\beta$          & LGBM-0.99           \\ \cline{2-3}
    & AChE           & XGB-0.98          \\ \hline
\end{tabular}
\end{table}

\section{Results}
\subsection{Testing and Validation}
\begin{figure}[H]
    \centering
    \caption{AUC-ROC for AChE Testing}
    \includegraphics[width=0.5\textwidth]{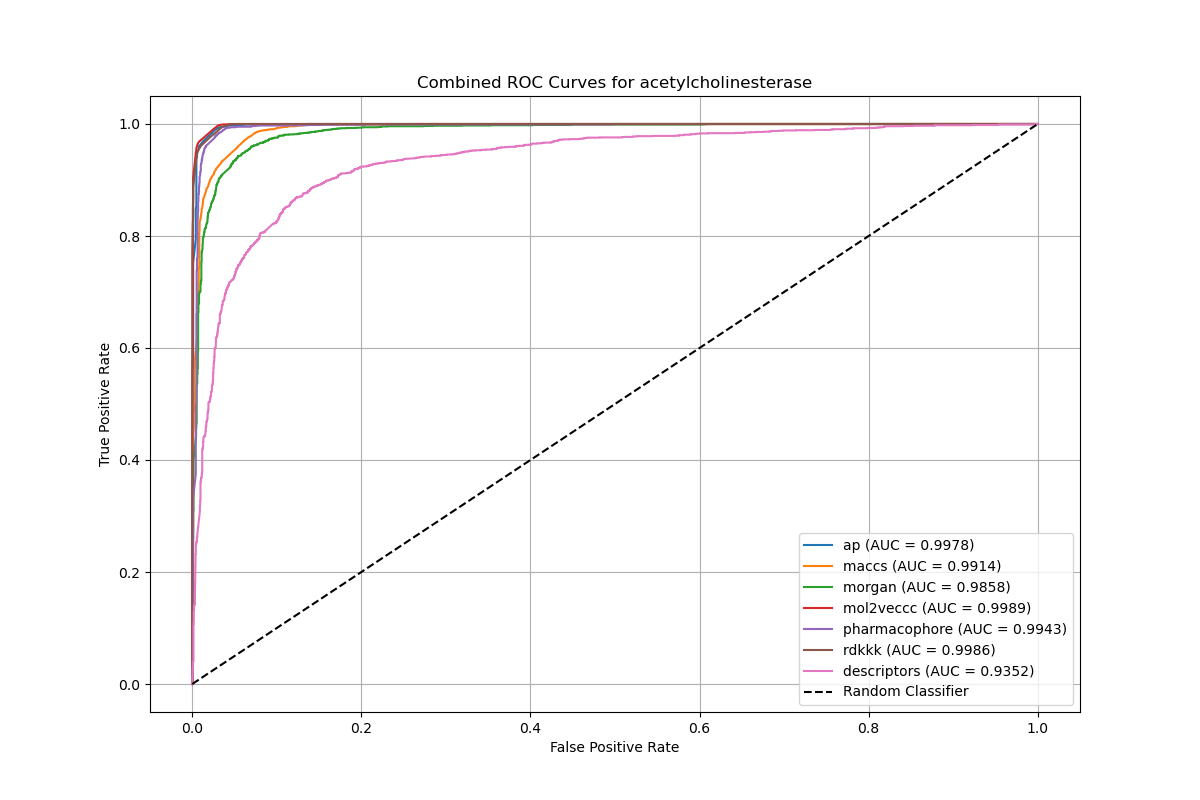}
    \label{fig:AChE_testing} 
\end{figure}

\begin{figure}[ht]
    \centering
    \caption{AUC-ROC for BACE-1 Testing}
    \includegraphics[width=0.5\textwidth]{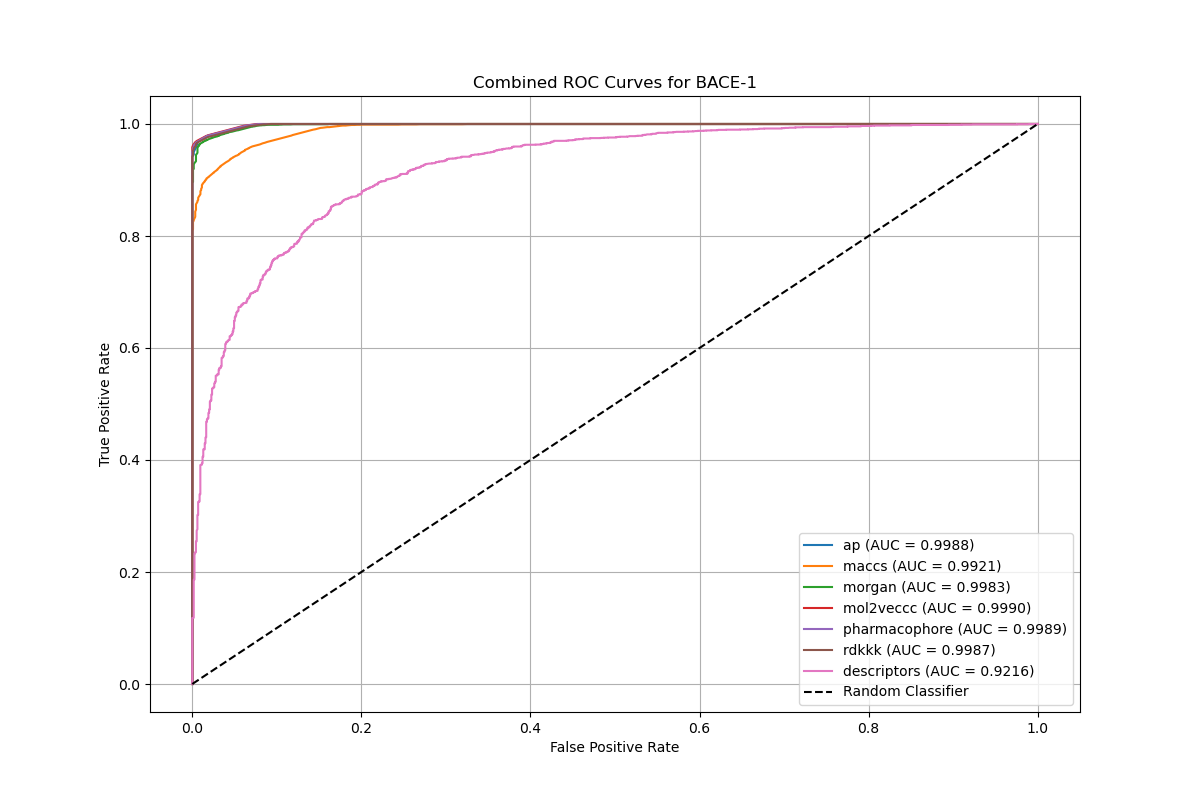}
    \label{fig:bace_testing} 
\end{figure}

\begin{figure}[ht]
    \centering
    \caption{AUC-ROC for GSK-3$\beta$ Testing}
    \includegraphics[width=0.5\textwidth]{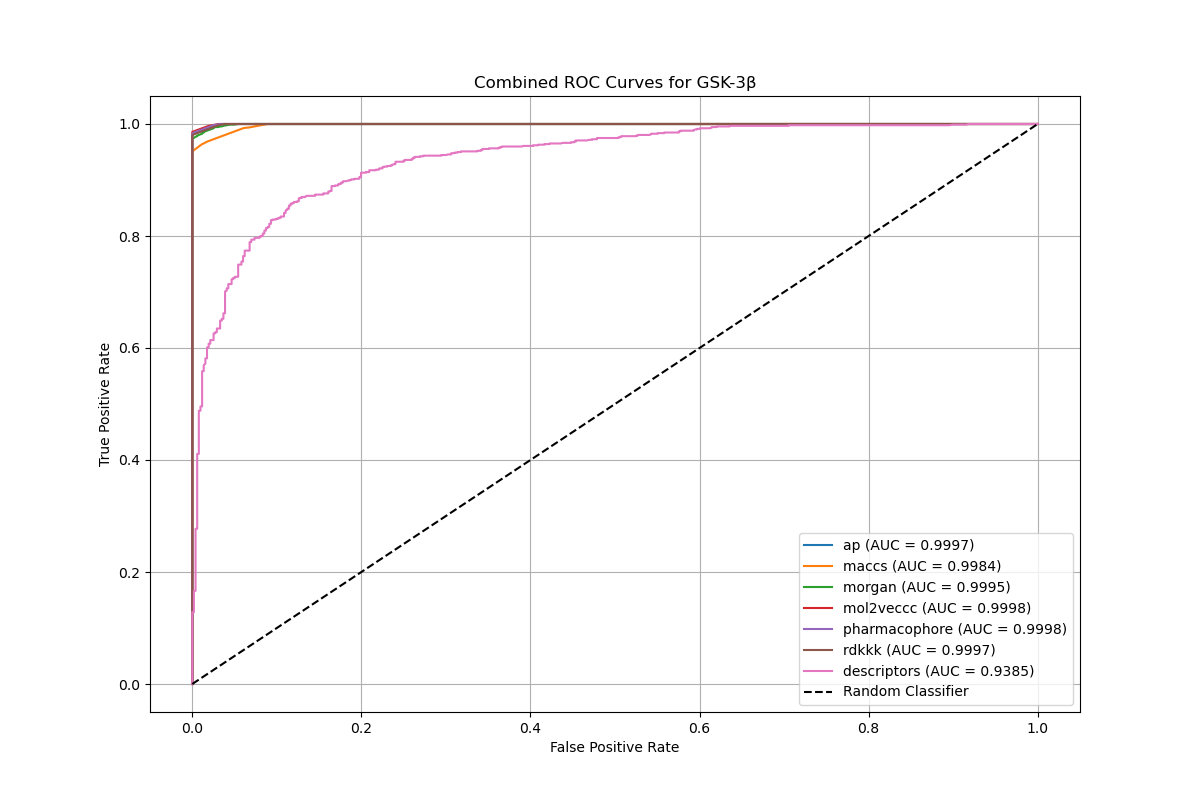}
    \label{fig:gsk_testing} 
\end{figure}

For all three targets, the accuracy, F1 Score, MCC, and AUC-ROC have been tabulated and compared with \cite{noviandy2023qsar}, \cite{Sandhu2022} and \cite{ijms242417233}. Only the best results out of all the fingerprints have been shown in table \ref{tab:testing_result}. For all our targets, Mol2Vec gave us the best metrics across all scores. The AUC-ROC curve for the same can be seen in figures \ref{fig:AChE_testing},\ref{fig:bace_testing},\ref{fig:gsk_testing}.

\begin{table}[ht]
\small
\centering
\caption{Result Comparison}
\label{tab:testing_result}
\begin{tabular}{|c|c|c|c|c|} 
\hline
\textbf{Model} & \textbf{Accuracy} & \textbf{F1 Score} & \textbf{AUC-ROC} & \textbf{MCC} \\ \hline
BACE-1 RF (paper) & 82.53 & 82.37 & NA & NA\\ \hline
BACE-1 (Ours) & 98.03 & 98.62 & 99.90 & 95.17 \\ \hline
AChE SVM (paper) & 85.86 & 83 & NA & 71 \\ \hline
AChE (Ours) & 98.19 & 97.96 & 99.89 & 96.35 \\ \hline
GSK-3$\beta$ KNN+RF (paper) & 71 & NA & NA & NA \\ \hline
GSK-3$\beta$ (Ours) & 99.09 & 99.29 & 99.98 & 98.04 \\ \hline
\end{tabular}
\end{table}


\begin{figure}[h]
    \centering
    \caption{AUC-ROC for AChE Validation}
    \includegraphics[width=0.5\textwidth]{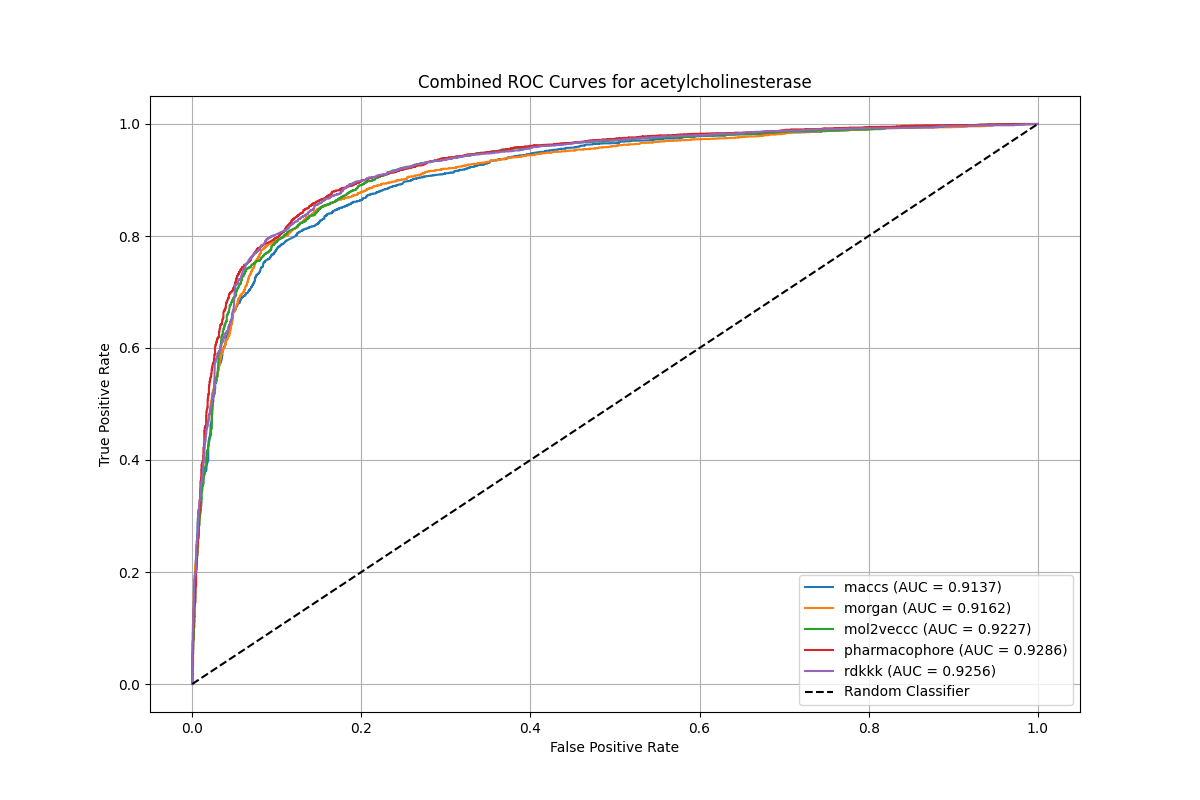}
    \label{fig:AChE_val} 
\end{figure}

\begin{figure}[h]
    \centering
    \caption{AUC-ROC for BACE-1 Validation}
    \includegraphics[width=0.5\textwidth]{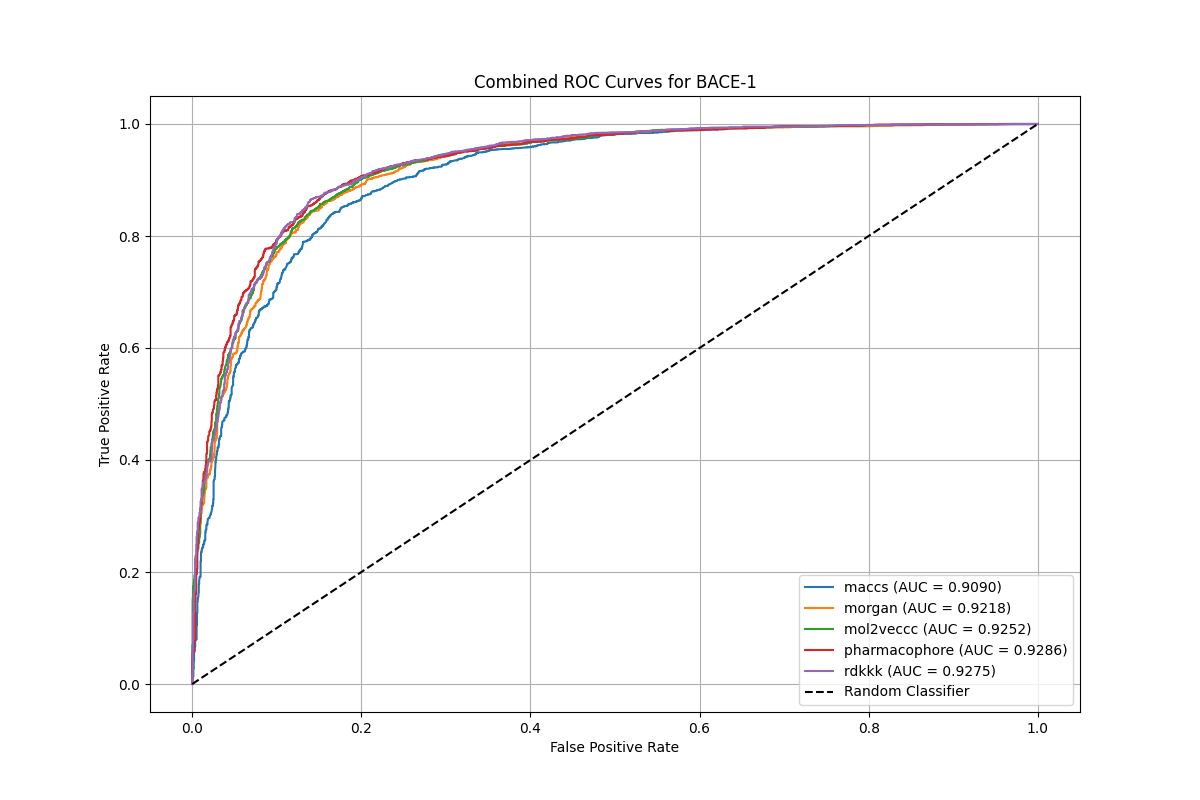} 
    \label{fig:bace_val} 
\end{figure}

\begin{figure}[h]
    \centering
    \caption{AUC-ROC for GSK-3$\beta$ Validation}
    \includegraphics[width=0.5\textwidth]{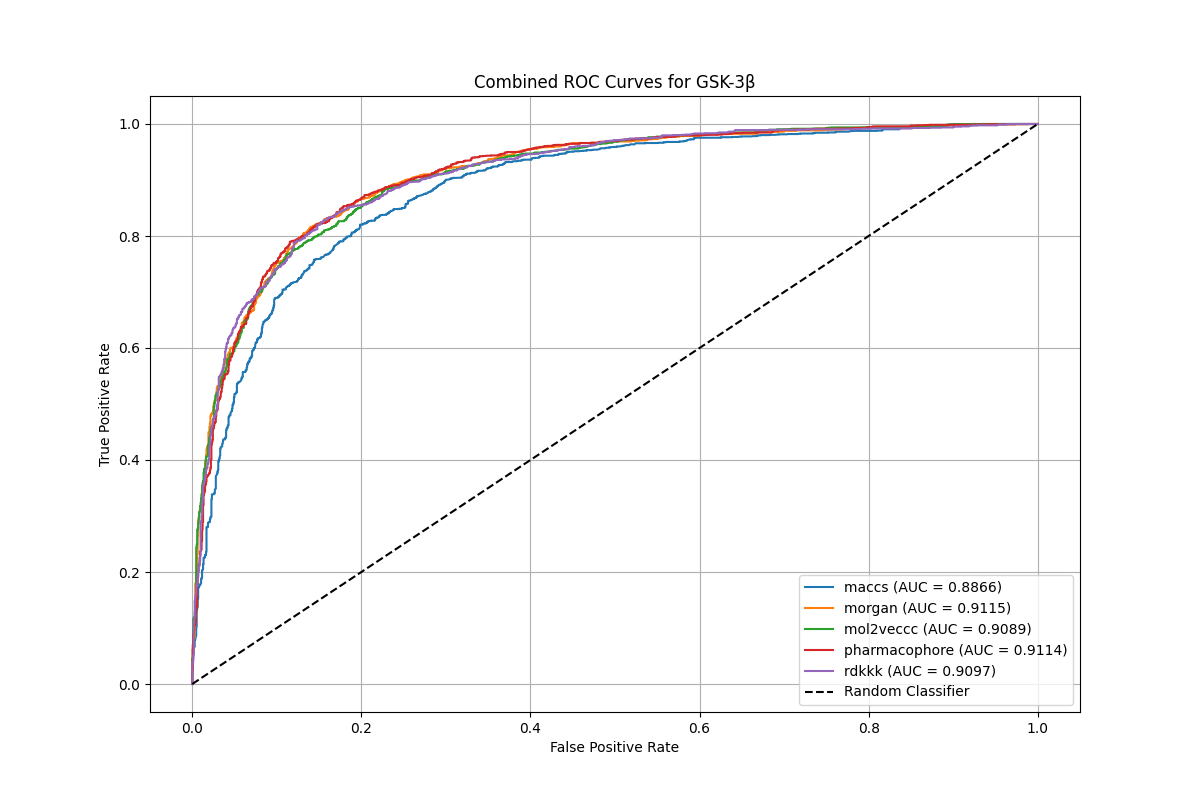}
    \label{fig:gsk_val} 
\end{figure}

Similarly, the models were run for our validation dataset from ChEMBL. The AUC-ROC for the same are in figures \ref{fig:AChE_val},\ref{fig:bace_val},\ref{fig:gsk_val}.
For validation, we didn't take atomic pair and molecular descriptors into account as they didn't generalize well and gave AUC-ROC scores in the range of 0.5-0.6. This shows that these descriptors may not be the best for developing machine learning models.


\subsection{Post-Validation Analysis}
\begin{figure}[h]
    \centering
    \caption{Top 5 Scaffolds for BACE-1}
    \includegraphics[width=0.5\textwidth]{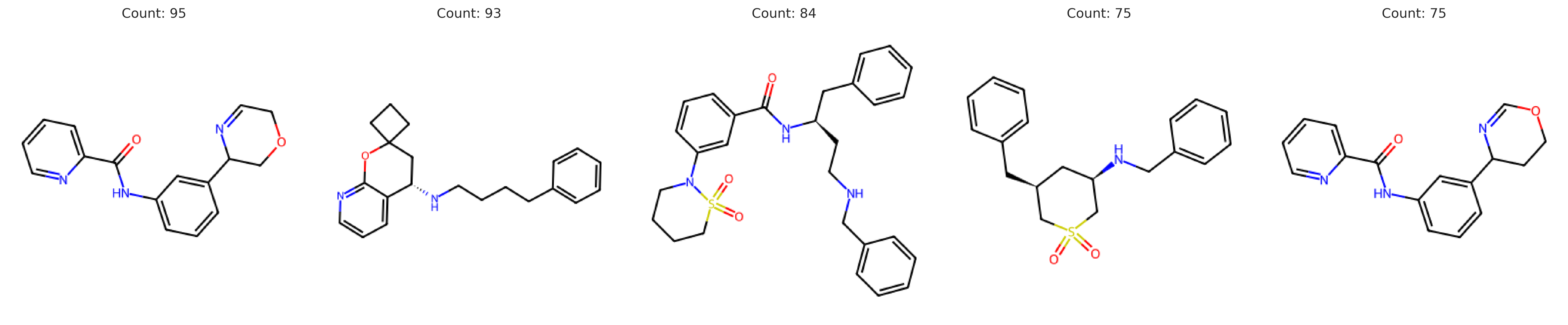}
    \label{fig:BACE1_scaffold} 
\end{figure}

\begin{figure}[h]
    \centering
    \caption{Chemical Space Distribution for BACE-1}
    \includegraphics[width=0.5\textwidth]{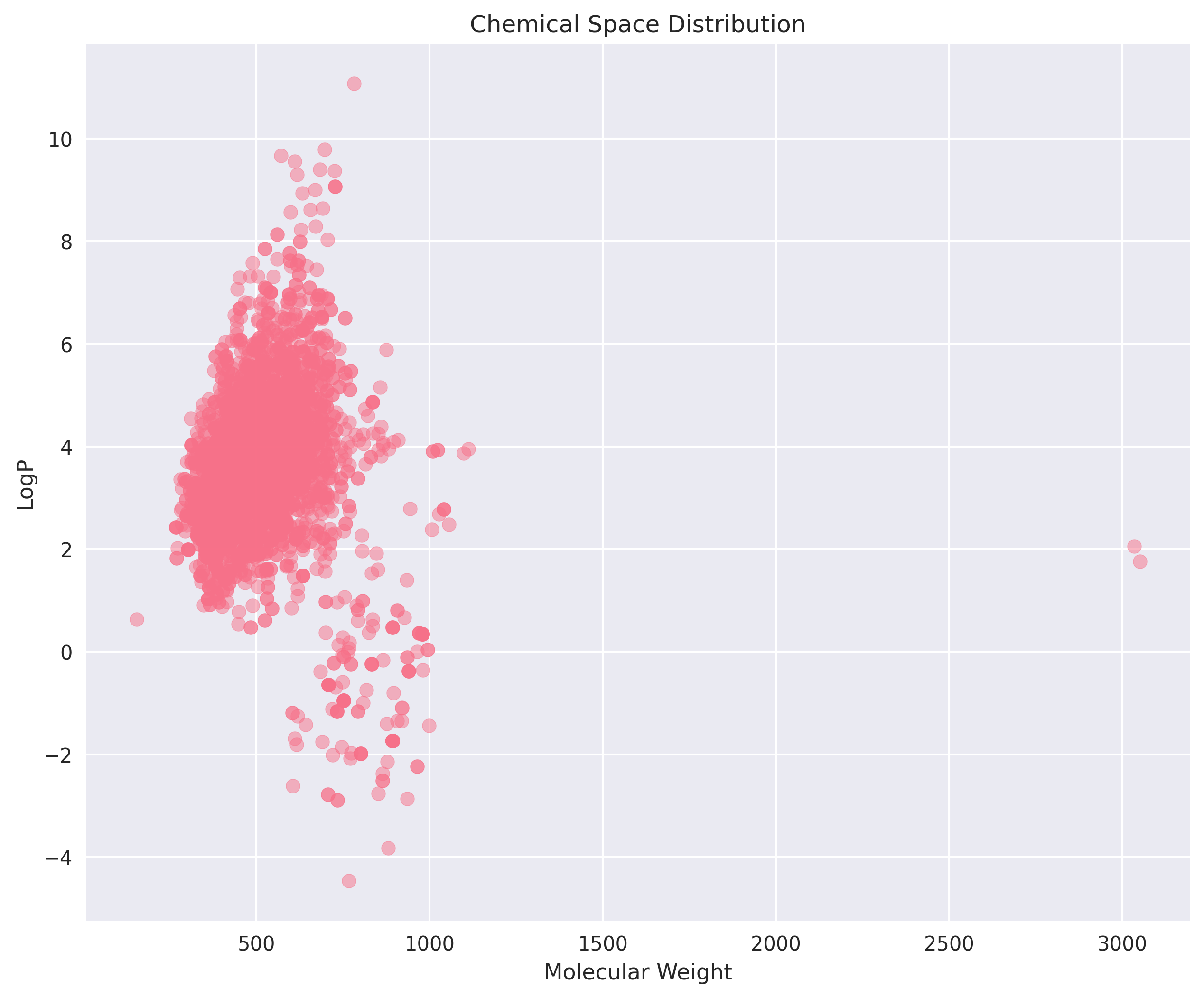}
    \label{fig:BACE1_space} 
\end{figure}

\begin{figure}[h]
    \centering
    \caption{Lipinski Violations for BACE-1}
    \includegraphics[width=0.5\textwidth]{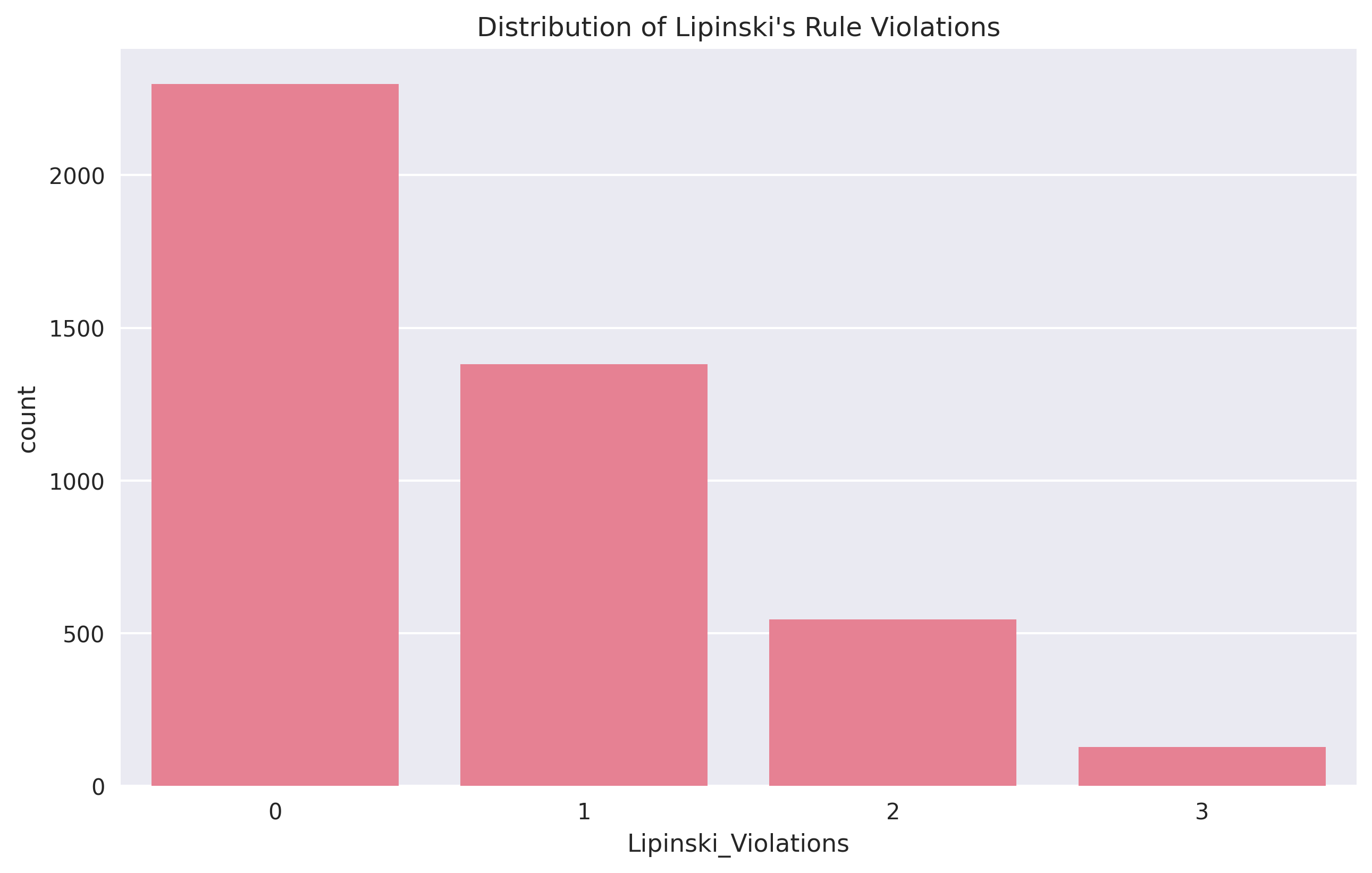}
    \label{fig:BACE1_lipinksi}
\end{figure}

\begin{figure}[h]
    \centering
    \caption{Property Analysis for BACE-1}
    \includegraphics[width=0.5\textwidth]{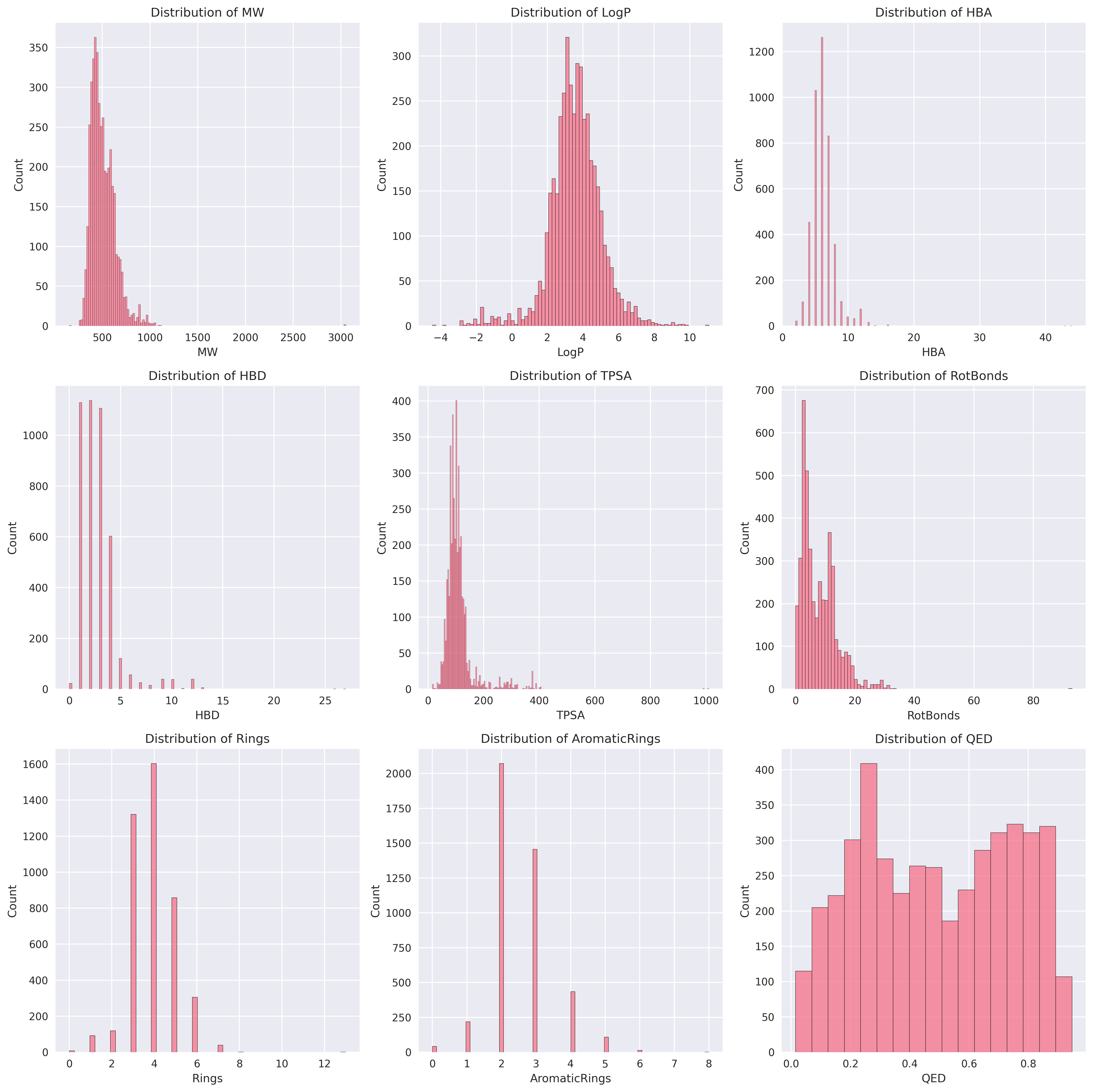}
    \label{fig:BACE1_prop}
\end{figure}

\begin{figure}[h]
    \centering
    \caption{Top 5 Scaffolds for AChE}
    \includegraphics[width=0.5\textwidth]{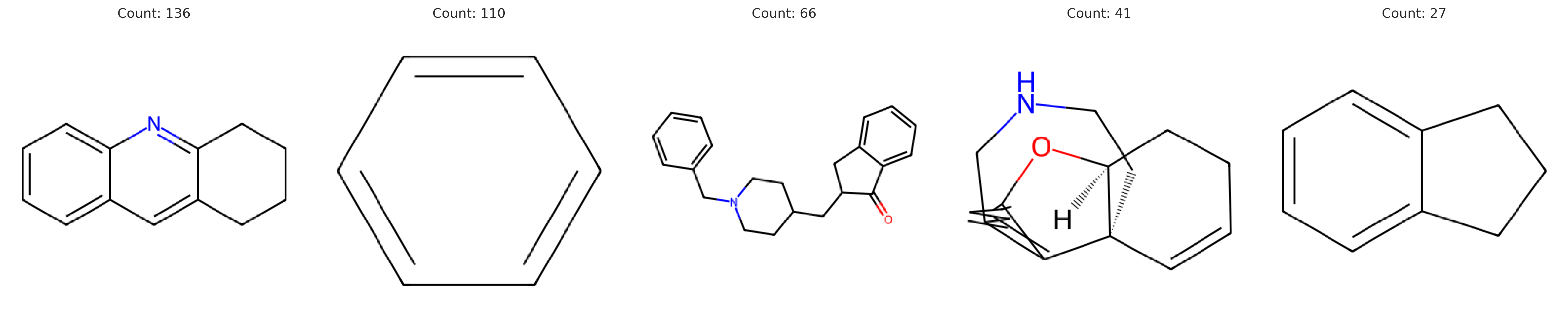}
    \label{fig:AChE_scaffold}
\end{figure}

\begin{figure}[h]
    \centering
    \caption{Chemical Space Distribution for AChE}
    \includegraphics[width=0.5\textwidth]{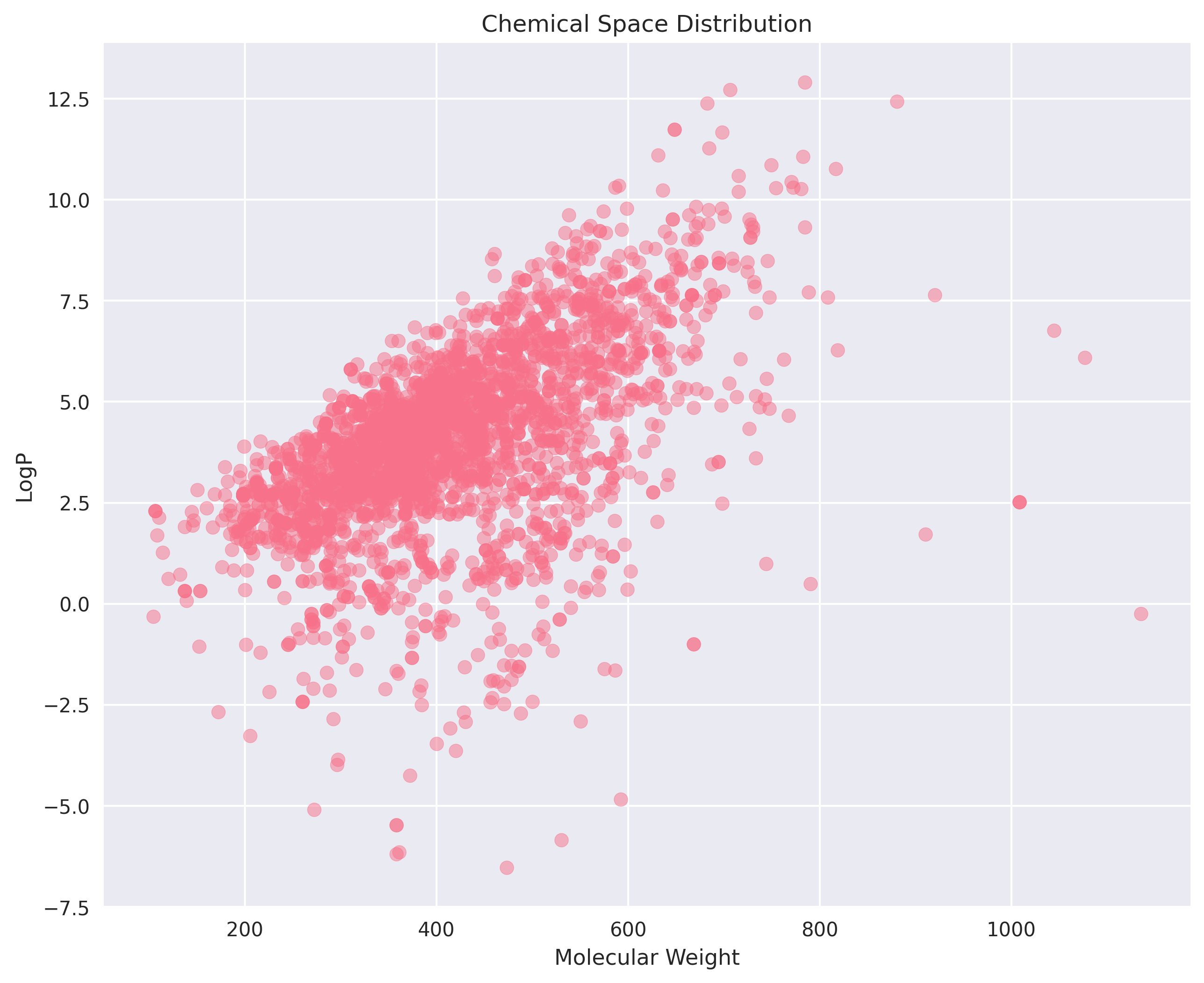}
    \label{fig:AChE_space} 
\end{figure}

\begin{figure}[h]
    \centering
    \caption{Lipinski Violations for AChE}
    \includegraphics[width=0.5\textwidth]{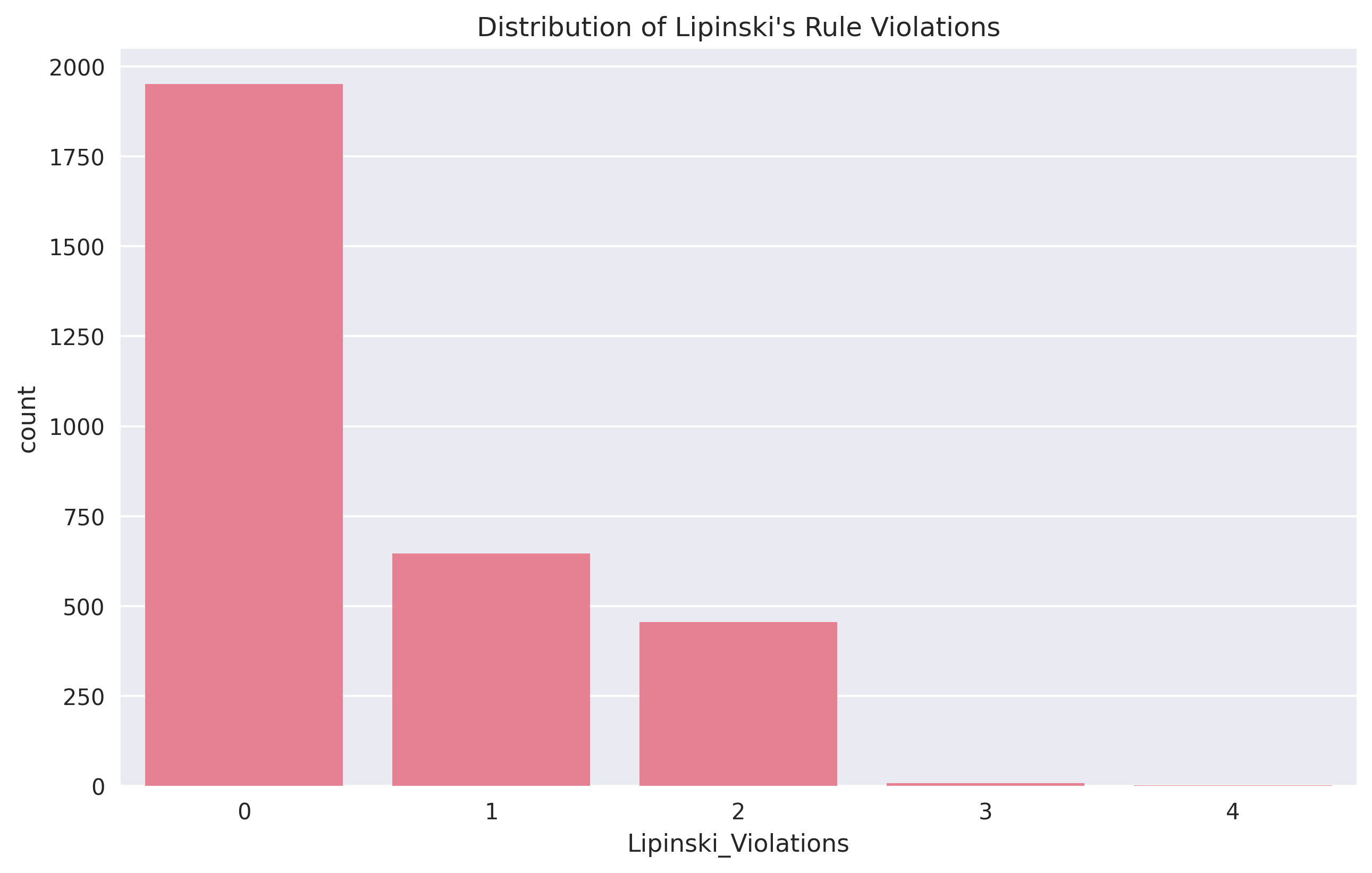}
    \label{fig:AChE_lipinksi}
\end{figure}

\begin{figure}[h]
    \centering
    \caption{Property Analysis for AChE}
    \includegraphics[width=0.5\textwidth]{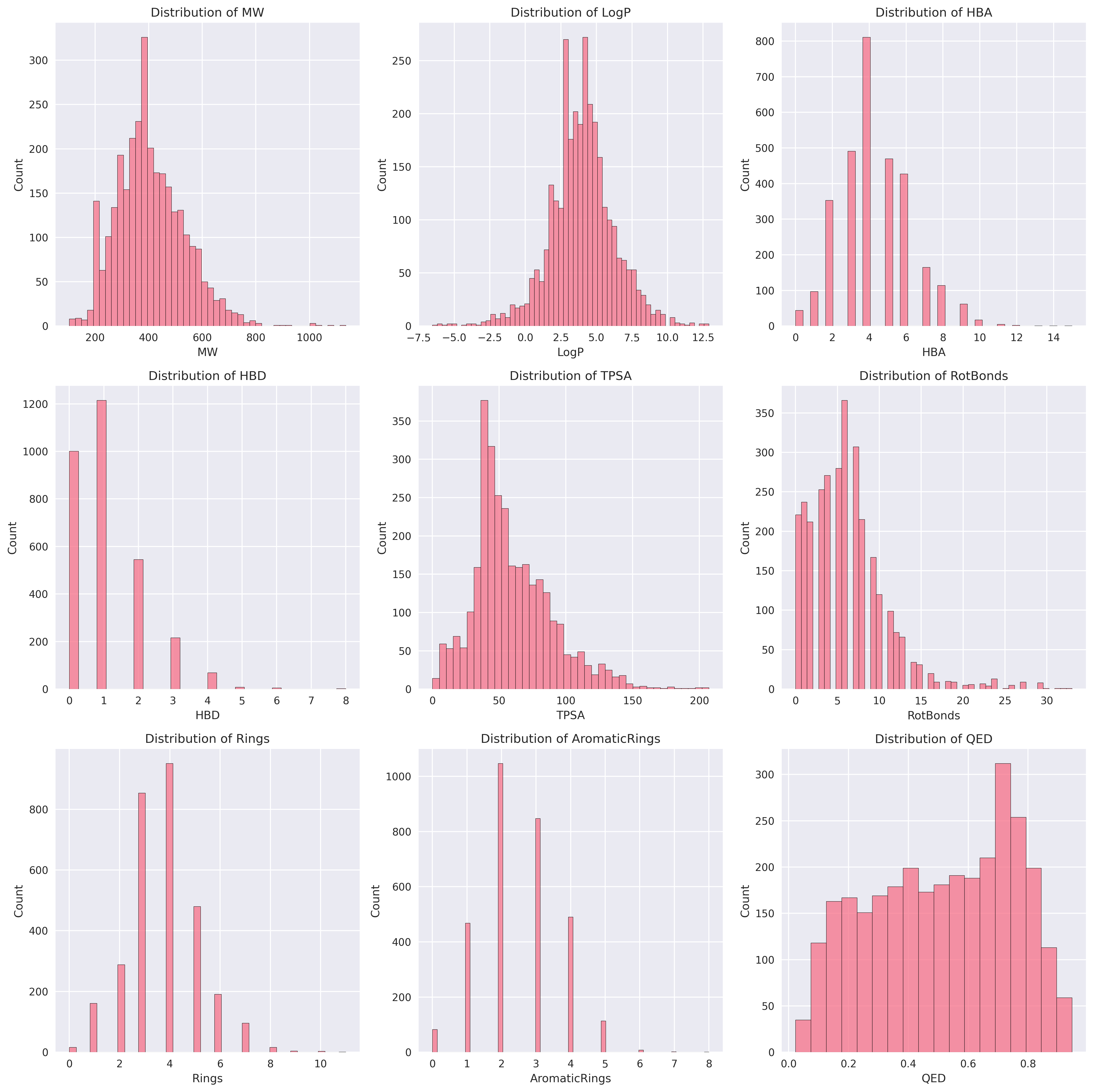}
    \label{fig:AChE_prop}
\end{figure}

\begin{figure}[h]
    \centering
    \caption{Top 5 Scaffolds for GSK-3$\beta$}
    \includegraphics[width=0.5\textwidth]{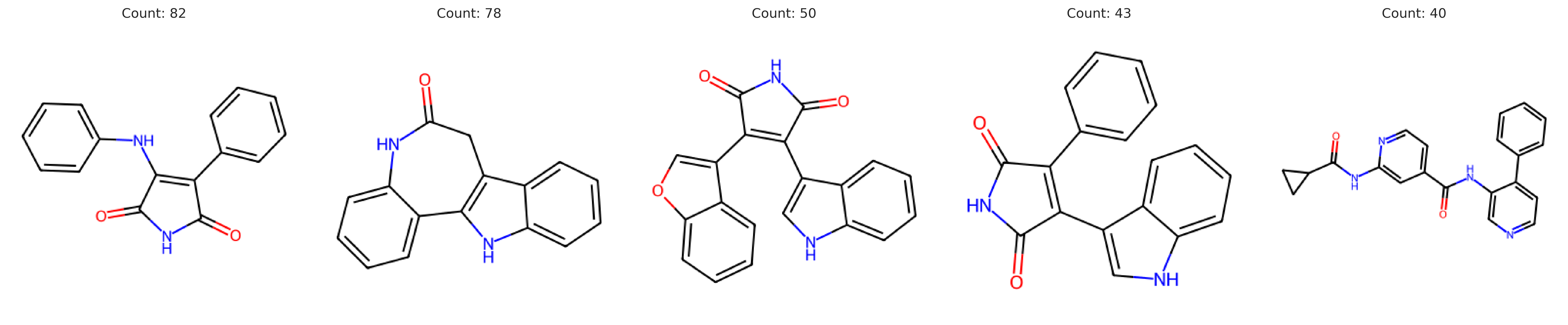}
    \label{fig:gsk_scaffold} 
\end{figure}

\begin{figure}[ht]
    \centering
    \caption{Chemical Space Distribution for GSK-3$\beta$}
    \includegraphics[width=0.5\textwidth]{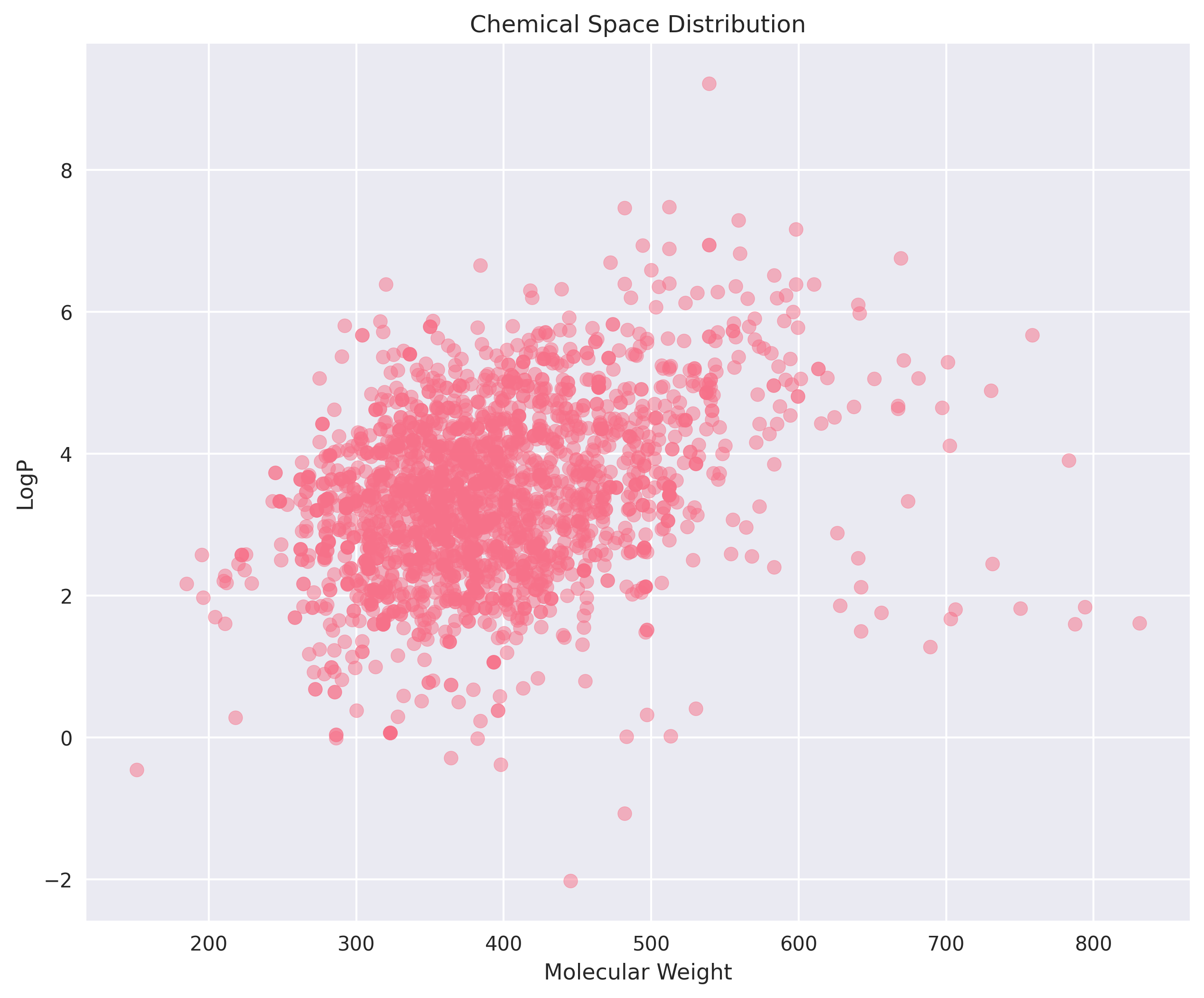}
    \label{fig:gsk_space} 
\end{figure}

\begin{figure}[ht]
    \centering
    \caption{Lipinski Violations for GSK-3$\beta$}
    \includegraphics[width=0.5\textwidth]{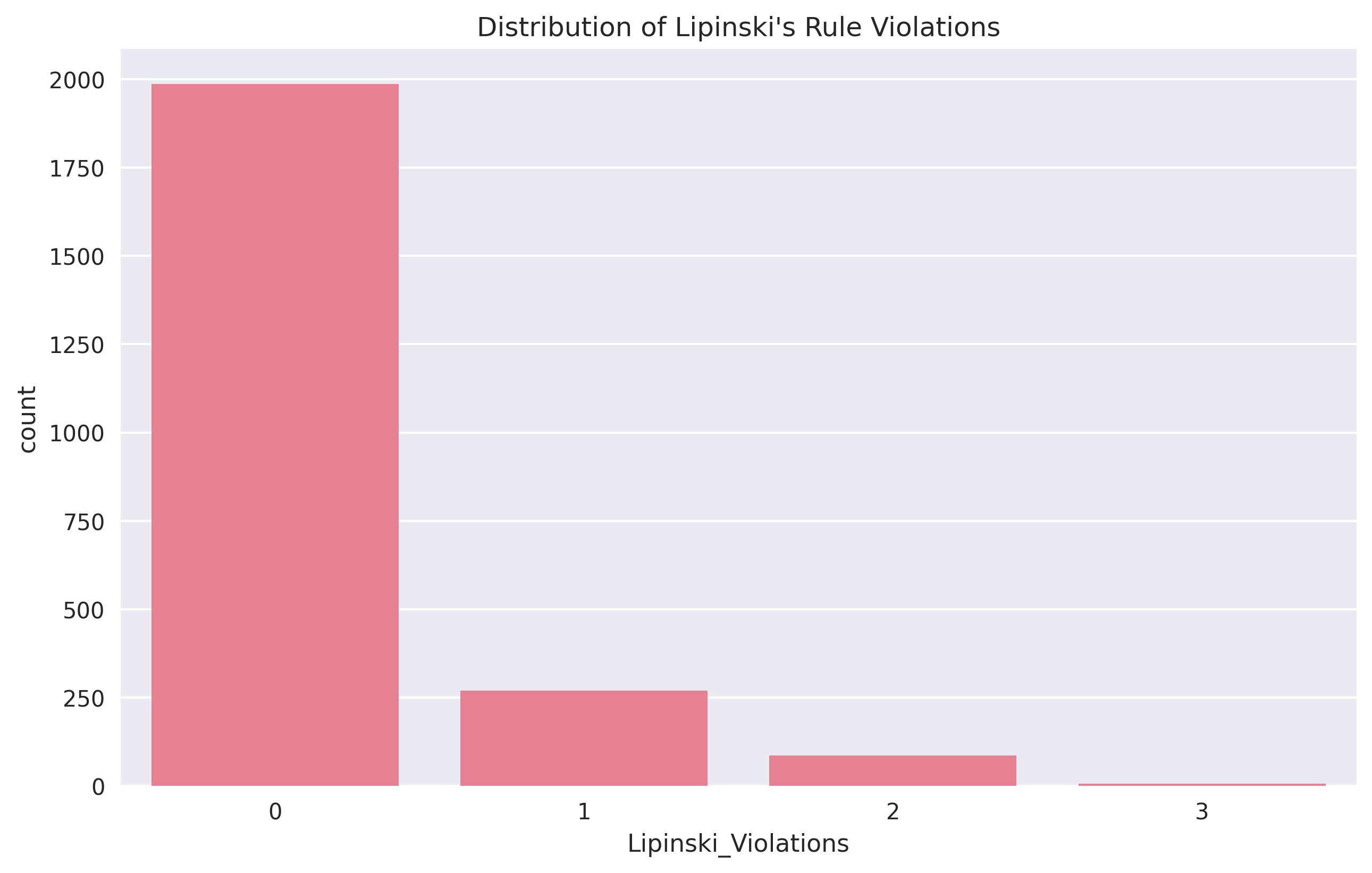}
    \label{fig:gsk_lipinksi}
\end{figure}

\begin{figure}[ht]
    \centering
    \caption{Property Analysis for GSK-3$\beta$}
    \includegraphics[width=0.5\textwidth]{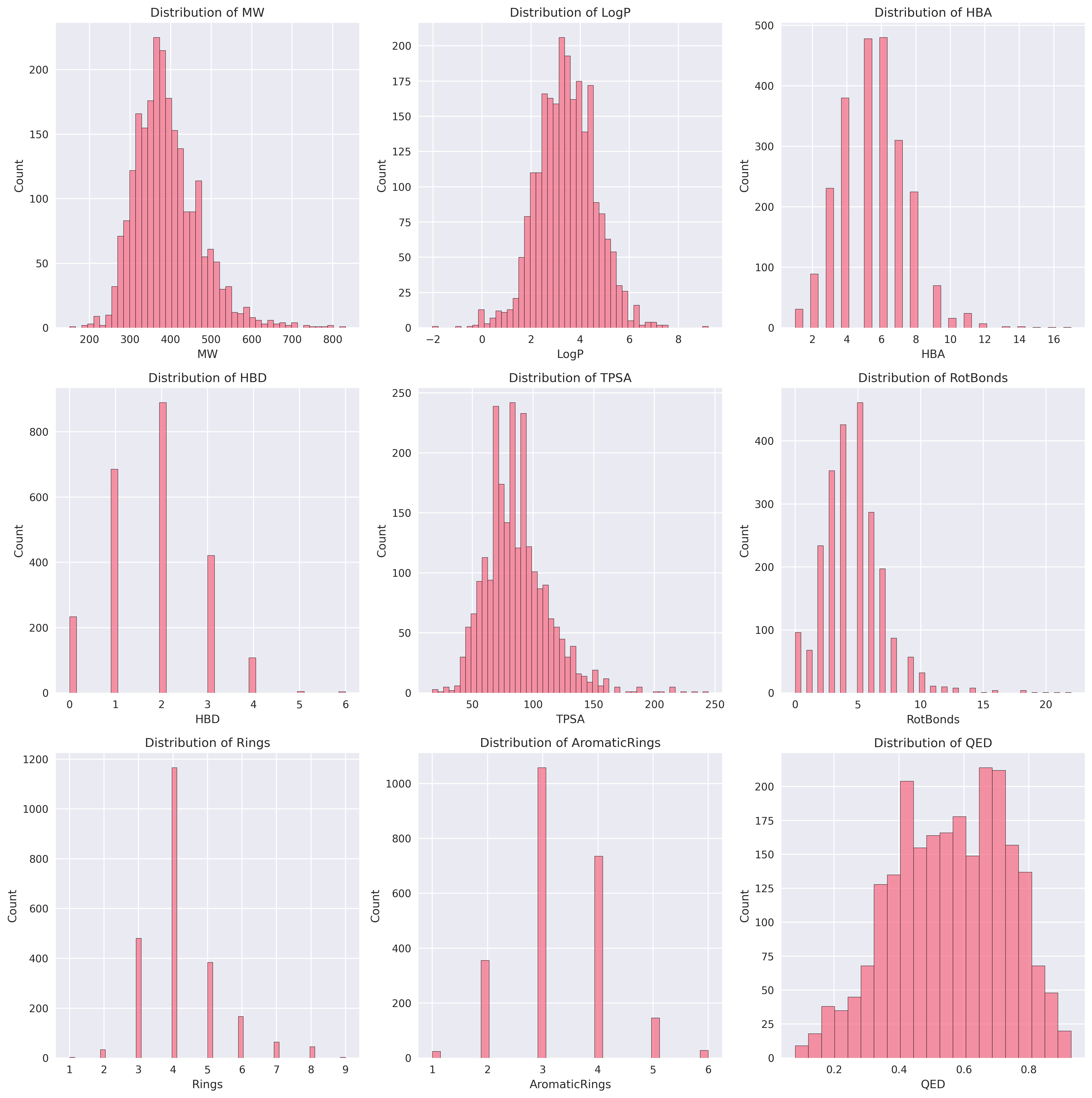}
    \label{fig:gsk_prop}
\end{figure}

After performing validation, we performed an in-depth analysis of the dataset for the best models for each target. These were done to explain the model's generalizability and alignment with real-world drug discovery requirements. We did the following analysis:
\begin{itemize}
    \item Chemical Space Distribution: This analysis visualizes the spread of chemical properties (e.g., molecular weight and LogP) in the dataset, helping us understand how well the validation data covers the broader chemical space. 
    \item Lipinski Violations: Lipinski's Rule of Five serves as a guideline for identifying drug-like molecules, particularly those suitable for oral administration. By analyzing violations, we can determine which molecules predicted by the model are more likely to succeed in preclinical development. Lesser the violations, the more viable the molecule.
    \item Scaffold Analysis: This also targets the generalizability of the model and highlights which structures the model can perform well on. Scaffold diversity is essential for identifying unique chemical cores that could serve as starting points for novel drug discovery efforts.
    \item Property distribution: Drug discovery workflows rely on molecules with optimal physicochemical properties. We can analyze these distributions to understand where the model may need refinement to align better with the requirements of viable drug candidates. 
\end{itemize}
For all our targets, the pharmacophore fingerprint gave the best results. For BACE-1, the results are summarised in figures \ref{fig:BACE1_scaffold},\ref{fig:BACE1_space},\ref{fig:BACE1_lipinksi},\ref{fig:BACE1_prop}, similarly in figures \ref{fig:AChE_scaffold},\ref{fig:AChE_space},\ref{fig:AChE_lipinksi},\ref{fig:AChE_prop} for AChE and figures \ref{fig:gsk_scaffold},\ref{fig:gsk_space},\ref{fig:gsk_lipinksi},\ref{fig:gsk_prop} for GSK-3$\beta$.




\section{Discussion and Future work}
We got many insights into our models from our post-validation analysis. They all performed exceptionally well for the validation and test dataset, crossing 0.9 for AUC-ROC in all cases. Two of our chosen fingerprints didn't do well and weren't included in the validation analysis. 

We shall consider the plots for each target:
\begin{itemize}
    \item BACE-1: Here, the best validation AUC-ROC was 92.86\% , with a validation accuracy of 86.63\%. The top 5 scaffolds and their counts account for only 422 entries, barely 6.5\% of the total data. However, the molecules are quite clustered (Figure \ref{fig:BACE1_space}), highlighting that the data doesn't
    have high diversity in physicochemical properties, which is also observed in Figure \ref{fig:BACE1_prop}. Considering this dataset, there are approximately 2500 molecules that violate no laws of Lipinski's, making them extremely viable for oral drug administration candidates. From this, we can conclude that the model performs well with high accuracy, but the dataset has limited chemical diversity, which may affect the generalizability of the model to novel, diverse compounds.
    \item AChE: The best validation AUC-ROC and accuracy were 92.86\% and 85.20\%, respectively. The top 5 scaffolds account for only 5\% of the total data. Here, we observe that the molecules aren't clustered in one corner, and there is more diversity observed considering their spread. Even from Figure \ref{fig:AChE_prop}, we can see that the values vary over a range and aren't entirely in a small range. The amount of molecules that can also be considered viable for drug candidates is around 2000. Considering this, we can say that the model would be able to perform comparatively better in terms of BACE-1 for unknown datasets, but the number of viable drug candidates is less. 
    \item GSK-3$\beta$: The best validation AUC-ROC and accuracy were 91.14\% and 82.93\% respectively. The top 5 scaffolds account for around 8\% of the entire data. The spread in molecules indicates a moderately rich diversity. The properties in Figure \ref{fig:gsk_prop} follow a similar trend to AChE. The number of viable molecules, however, is higher in ratio here to all the previous targets. 
\end{itemize}

From these, we can understand that our models have a moderately good generalizability capability. It performed the best for BACE-1 just considering the metrics. However, taking other aspects into account, AChE models would perform better for unknown data. 

This may be something that would vary with the data collected, but the number of viable molecules for inhibition of GSK-3$\beta$ wrt to the total molecules is significantly higher than the other targets. This suggests that this target might be more promising for future work.

For future efforts, the next step is to collate a larger and more diverse dataset. The current pipeline performs exceptionally well, which indicates that it is capturing key features effectively. Expanding to a larger dataset will help us further refine our models. Additionally, a key goal is to consolidate the existing models into a unified framework that can handle multi-class classification, transitioning from binary classification. Once we identify promising compounds, we can proceed with docking analysis and clinical trials to assess these drugs' real-world viability and confirm our models' continued efficacy in practical applications.

\section{Conclusion}
From our rigorous testing, we can conclude as an initial step that models like XGBoost and LightGBM are good baseline models for classification. However, we worked with binary classification and the efficacy of the models can't be commented on when it is extended to multi-class classification. Additionally, while our models demonstrate strong performance on the testing and validation datasets, the generalizability to unseen or more diverse data remains an open question. Future improvements can be achieved by incorporating more diverse and larger datasets to enhance model robustness. Finally, exploring more advanced methods, such as transfer learning or ensemble techniques, may offer further enhancements to our predictive capabilities in drug discovery.

\printbibliography[title={7. References}]

\end{document}